\documentclass [12pt,preprint]{aastex}
\usepackage{epsfig}
\usepackage{natbib}

\begin{document}

\voffset-0.5cm
\newcommand{\gsim}{\hbox{\rlap{$^>$}$_\sim$}}
\newcommand{\lsim}{\hbox{\rlap{$^<$}$_\sim$}}

\title{Super Luminous Supernova and Gamma Ray Bursts}

\author{Shlomo Dado\altaffilmark{1} and Arnon Dar\altaffilmark{1}}

\altaffiltext{1}{Physics Department, Technion, Haifa 32000, Israel}
\begin{abstract}

We use a simple analytical model to derive a closed form expression for 
the bolometric light-curve of super-luminus supernovae (SLSNe) powered by 
a plastic collision between the fast ejecta from core collapse supernovae 
(SNe) of types Ib/c and IIn and slower massive circum-stellar shells, 
ejected during the late stage of the life of their progenitor stars preceding 
the SN explosion. We demonstrate that this expression reproduces well the 
bolometric luminosity of SLSNe with and without an observed gamma ray 
burst (GRB), and requires only a modest amount ($M < 0.1\,M_\odot$) of 
radioactive $^{56}$Ni synthesized in the SN explosion in order to explain 
their late-time luminosity. Long duration GRBs can be produced 
by ordinary SNe of type Ic rather than by  'hypernovae' -
a subclass of superenergetic SNeIb/c.

\end{abstract}

\keywords{gamma-ray burst: general, supernovae: general}

\maketitle

\section{Introduction}

The progenitor stars of core-collapse supernovae of type Ib/c, are 
stripped of their envelope through strong winds and/or major eruptions 
during the final stages of their life before their explosion (e.g., 
Pastorello et al.~2010; Quimby et al.~2011, and references therein). The 
ejected massive shells in these eruptions sweep up slower stellar winds, 
which were blown before these eruptions, and create a matter-clean space 
surrounding the progenitor star.  The interaction of the radiation from 
the SN explosion with such slowly moving  circumstellar (CS) 
shells  often produces delayed emission of narrow lines, which stops when 
the SN  ejecta collide with the CS shell (Chugai~1990,1992).

Light from the progenitor star back-scattered by  CS shell(s) into the 
matter-free space around the progenitor star produces a glory - a halo of 
scattered light surrounding the progenitor star. In the cannonball 
(CB) model of gamma ray bursts (GRBs), long duration GRBs are produced by 
inverse Compton scattering of glory photons by the electrons in the highly 
relativistic bipolar jets of plasmoids (cannonballs) of ordinary matter, 
which presumably are ejected in mass accretion episodes of fall-back material 
on the newly formed central object (neutron star or black hole) in stripped-envelope 
supernova explosions (e.g., Dar \&  
R\'ujula~2000,2004; Dado et al.~ 2009). In this scenario, SN explosions 
that produce long GRBs (SNe-GRB) are ordinary core-collapse 
SNe of type Ic where the kinetic energy of the ejecta is typically a few 
$10^{51}$ ergs, rather than 'hypernovae' - hypothetical super energetic 
core collapse SNIc explosions, where the kinetic 
energy of the ejecta exceeds a few $10^{52}$ ergs and their bolometric 
light-curve is powered by the radioactive decay of $M\gg 0.1\, M_\odot$ of 
$^{56}$Ni synthesized in the explosion (e.g., Iwamoto et al.~1998; 
Nakamura et al.~2001). So far, the CB model has been very successful in 
predicting/reproducing the main observed properties of long duration GRBs 
(e.g., their rate, location in star formation regions, association with 
core collapse SNe of type Ib/c, typical photon energy, multi-pulse 
structure, pulse shape and duration, spectral evolution of the individual 
pulses and large photon polarization) and the observed correlations 
between them (see, e.g., Dado et al.~2009; Dado \& Dar 2012, and 
references therein).

The CB model scenario implies that in SNeIc the fast SN ejecta may collide 
with a slowly expanding massive CS shell ejeted some time 
before the explosion. Such a collision may produce a very luminous SNIc 
and even super-luminous (SLSN), which are powered mainly by the 
collision rather than by a large mass of $^{56}$Ni synthesized in the 
explosion. Because of relativistic beaming, most of the GRBs that are 
produced in SNeIc and SLSNe are beamed away from Earth and are not 
observed. Indeed, most SNeIc and SLSNe are not accompanied by an observed 
GRB. The first discovered SLSN without an associated GRB was SN1999as 
(Knop et al.~1999) at a redshift z=0.127, which was much more luminous 
than the very bright SNe of type Ib/c that produced observed GRBs such as 
SN1998bw (Galama et al.~1998), that produced GRB 980425 (Soffitta et 
al.~1998, Pian et al.~2000), SN2003dh (Stanek et al.~2003; Hjorth et 
al.~2003) that produced GRB 030329 (Vanderspek et al.~2003), and SN2006aj 
(Campana et al.~2006; Pian et al.~2006; Sollerman et al.~2006) that 
produced GRB 060218 (Cusumano et al.~2006). More recently, transient 
surveys that were monitoring many square degrees of the sky every few 
nights have discovered several additional SLSNe in the nearby Universe 
without an observed GRB. The first one was SN2005ap (Quimby et al.~2007). 
The absence of hydrogen in its spectrum, its very broad lines, and its 
energetics led Quimby et al.~ (2007) to propose that SN2005ap could have 
been produced by the same mechanism that produces SNe with an observable 
GRB. Other discoveries of SLSNe without an observed GRB include SN2003ma 
behind the Large Magellanic Cloud at z =0.289 by the SuperMACHO 
microlensing survey (Rest et al.~2011), SN2006gy (Smith et al.~2008,2010), 
SN2007bi (Gal-Yam et al.~ 2009), SN2008am (Chatzopoulos et al.~2011), 
SN2010hy (Kodros et al.~ 2010; Vinko et al.~2010), SN2010gx (Pastorello et 
al.~2010), SN 2010jl (Stoll et al.~2011), and SCP 06F6 (Quimby et 
al.~2011).

Alternative mechanisms which were invoked in order to explain the 
observed luminosity of SLSNe include:\\
I. Radioactive decay of large amounts 
(several $M_\odot$) of radioactive $^{56}$Ni produced in pair-instability 
explosions of extremely massive stars (Rakavy \& Shaviv~1967; Barkat et 
al.~1967; Heger \& Woosley~2002; Waldman~2008; Gal-Yam et al.~2009; 
Yoshida \& Umeda~2011), which are efficiently mixed in the SN ejecta. \\ 
II. Efficient conversion of kinetic energy of the SN ejecta into thermal 
energy in SN explosions inside optically thick winds (Falk \& 
Arnett~1973,1977; Ofek et al.~2007; Smith \& McCray~2007; Smith et 
al.~2010; Balberg \& Loeb~2011; Chevalier \& Irwin~ 2011,2012; Moriya et 
al.~2012; Chatzopoulos et al.~2012; Ginzburg \& Balberg~2012; Ofek et 
al.~2012).\\
III. Collision(s) of the fast SN ejecta with slowly expanding 
dense circum-stellar shell(s) ejected by the progenitor star 
sometime before the SN explosion (Grassberg et al.~1971;
Moriya et al.~2012), 
supplemented by energy release in the radioactive decay chain\\
${\rm^{56}Ni\rightarrow^{56}Co\rightarrow^{56}Fe}$ 
of  $M(^{56}{\rm Ni}) \ll M_\odot$  synthesized in the explosion.

The rapid decay of the bolometric light-curve of SLSNe such as SN2010gx, 
and the very large mass, $\sim 10\, M_\odot$, of $^{56}$Ni needed to 
explain its peak luminosity, however, indicate that the pair instability 
mechanism where a large mass of $^{56}$Ni is produced (scenario I) is 
unlikely to be its power source (Pastorello et al.~2010). In scenario II, 
the progenitor star explodes into a dense wind, and the strong 
shock that presumably explodes it breaks out into the wind. This strong 
shock is assumed to convert the kinetic energy which it imparts to the 
ejecta in SN explosions into internal thermal energy of the wind. 
Numerical simulations of core-collapse SNe, however, so far  have not 
produced consistently strong enough shocks that can reproduce 
the observed SNe where the typical kinetic energy of the debris is a few 
$10^{51}$ ergs. But, by adjusting the wind parameters and the energy 
deposited in it, and by introducing many simplifying 
assumptions, 
Ginzburg and Balberg (2012) were able to calculate bolometric light curves 
for some SLSNe, which look like those observed. Scenario III has not been 
studied yet in detail with numerical hydrodynamical codes (Ginzburg \& 
Balberg 2012). However, scenario III is strongly suggested by observations 
of SNn of type Ib/c and by the success of the CB model of GRBs in 
predicting the main observed properties of long GRBs produced in SNe of 
Type Ic.

In this letter we use a simple analytical model based on only a few general 
assumptions to derive a closed form expression for the bolometric 
luminosity of SLSNe in scenario III. It involves only few adjustable 
parameters. We use it  to demonstrate that collisions between 
the fast ejecta from core collapse SNe of types Ib/c and their massive 
circum-stellar shells, which were ejected in eruptions of their massive 
progenitors in the years preceeding their SN explosion, together with a 
modest amount of radioactive isotopes, which were synthesized in 
the SN explosion and deposited in the ejecta, can reproduce quite well the 
bolometric light-curves of both the supernovae that were observed in 
association with GRBs  and those of the recently 
discovered SLSNe without an observed GRB. We conclude with a short 
discussion of the implications for SN explosions and GRBs.

\section{Collision of the SN ejecta with a dense CS shell}

The observed narrow lines in SNeIc indicate  that the dense 
circum stellar shells (CSS) that emit them  are expanding with a 
velocity of a few hundreds km/s at a typical distance of $R\sim 
3\times 10^{16 \pm 1}$ cm from the exploding star, and have a typical 
baryon density  $n\sim 10^{8 \pm 2}\, {\rm cm^{-3}}$ and a typical mass 
$M_{css}\sim 1-10\, M_\odot$.  The SN ejecta of mass $M_{ej}$ that have a 
much faster velocity, $v_{ej}\sim 10^4\, {\rm km\, s^{-1}}$, overtake the 
slower massive CSS typically  within $R/(v_{ej}-v_{css})\sim 
10^{1.5\pm 0.5}$ days. 

We assume that the collision between the SN ejecta and a  CS shell
is a plastic collision.  Then, energy - momentum conservation implies that 
the center of mass (CM) energy 
$\mu\, v^2/2$ is converted to internal energy, where $\mu=M_{ej}\, 
M_{css}/(M_{ej}+M_{css})$ is the reduced mass of the colliding shells and 
$v=v_{ej}-v_{css}\approx v_{ej}$ is their relative velocity. This internal 
energy is roughly a fraction $M_{css}/(M_{ej}+M_{css})$ of the kinetic 
energy of the incident ejecta, typically a few $10^{51}$ ergs. 
For a patchy CS shell, or a clumpy SN ejecta, that covers a solid angle 
$\eta\, 4\,\pi$, 
$M_{ej}$  must be replaced  by $\eta\, M_{ej}$. The thermal expansion 
speeds of the SN and CS shells are negligible compared to their 
radial velocities and their merger is completed within  relatively a short 
time.

Consider now the collision in the rest frame of the CS shell.  Most of the kinetic energy of the 
SN ejecta is carried by atomic nuclei of mass $\sim A\, m_p$ and charge $Z\, e$. For 
$\beta=v/c\sim 1/30$, their typical kinetic energies is $A\, m_p\,\beta^2\, c^2/2\sim A/2 $ MeV. 
In SNeIc, the CS shells consists mainly of the hydrogen and helium layers, which were stripped 
off from the progenitor star sometime before the SN explosion. The SN ejecta from such a massive 
star stripped off of its hydrogen and helium layers, consists mainly of nuclei heavier than 
helium nuclei.  Such nuclei lose their kinetic energy in the CS shell mainly by Coulomb 
interactions with electrons (ionization and scattering off free electrons) at a rate which is 
given approximately by the Bethe-Bloch energy loss rate of sub-relativistic nuclei in matter. For 
ejecta with velocity $v\sim 10^4$ km/s, $dE/dx\sim 1.35\, {\rm MeV\, Z^2\,\beta^{-2}\, cm^2\, 
g^{-1}}$, they lose their energy in the CS shell within $\leq 10^{-4}\, {\rm g\, cm^{-2}}$, 
whereas the typical column density of CS shells is a few ${\rm g\, cm^{-2}}$. Consequently, the 
kinetic energy deposition is mainly near the interface of the colliding shells, and it lasts 
relatively a very short time. This energy is converted almost instantaneously to thermal (black 
body) radiation ($a\, T^4 \gg n\,k\,T$) by bremsstrahlung and multiple scattering of the knocked 
on electrons, and is transported from the collision zone to the entire CS shell by diffusion and 
shock wave. This thermal radiation escapes into the interstellar medium (ISM) from a depth with 
an opacity $\tau_{opt}\sim 1$. If the opacity of the entire ionized shell is 
$\tau_{opt}=N_e\,\sigma_{_T}>1$ where $N_e$ is the electron column density of the shell, and 
$\sigma_{_T}\approx 0.67\times 10^{-24}\, {\rm cm^2}$ is the Thomson cross section for Compton 
scattering, then the photons that the thermal radiation from the collision cross the CS shell by 
'random walk' in a typical time $t_{r}\approx \Delta R^2/\lambda\, c = \tau_{opt}\,\Delta R/c $ 
where $\Delta R$ is the width of the CS shell whose density becomes nearly uniform for $R \gg 
\Delta R $.

The radioactive (ra)
decay chain ${\rm^{56}Ni\rightarrow^{56}Co\rightarrow^{56}Fe}$ 
in the collision zone  releases there additional energy  at a rate,    
\begin{equation}
\dot{E}_{ra}=7.76\times 10^{43}{M(^{56}{\rm Ni})\over M_\odot}\, 
    [e^{-t/8.77\,{\rm d}}+0.227\, e^{-t/111.4\,{\rm d}}]\,  {\rm erg\, 
s^{-1}}\,, 
\label{radecay}
\end{equation}  
For $t> 14$ d, the decay of $^{56}$Co 
dominates the radioactive energy release.

\section{The bolometric light-curve}
Let  $t=0$ be the explosion time and 
$t_c$ be  the collision time after the explosion.
Let us approximate the behaviour of  the CS shell 
by that of a container in thermal 
equilibrium  with a  black body radiation of  temperature $T$,
energy density $u=a\, T^4$, and  pressure $p=u/3$, 
whose radius $R$ expands at a constant rate 
$V=(m_{ej}\, v_{ej}+m_{css}\,v_{css})/(m_{ej}+m_{css})$
but whose width $\Delta R$ remains nearly constant.
Energy conservation implies that its  photospheric 
luminosity $L=4\, \pi\, R^2\, \sigma \, T^4/\tau_{opt}$ 
satisfies approximately 
\begin{equation}
t_{opt}\,\dot{L}+L=\dot{E}_{ra}\,, 
\label{Lum}
\end{equation}
where $t_{opt}= 4 \pi\, \Delta R\, \tau_{opt}/c$  as long as it 
is opaque ($\tau_{opt}>1$) and cooling by
adiabatic expansion of the CS shell was neglected compared to  
radiative 
cooling.  If the CS shell is patchy and covers only
a solid  angle $ \eta\, 4\,\pi$ where $\eta <1$, 
then the bolometric luminosity  $L= 4\,\pi\,\eta\, R^2\, \sigma \, 
T^4$ still satisfies Eq.(2) and 
yields a photospheric effective radius smaller than the true radius 
by a factor $\eta^{1/2}$.

The general solution of Eq.(2) for $t>t_c+t_{rw}$ is given by $L=L_c+L_p$
where $L_c(t)= L_c(t-t_c-t_{rw})\,  e^{-(t-t_c-t_{rw}))/\tau}$ 
is the general solution of 
the homogeneous part $\dot{L}+\,L/\tau_{opt}=0$ of Eq. (2), and 
$L_p(t)$ is a  particular solution of the entire equation, 
\begin{equation} 
L_p(t)= 7.76\times 
10^{43}{M(^{56}Ni)\over M_\odot}\,\left[
{e^{-t/\tau_{_{Ni}}}\over (1-\tau/\tau_{_{Ni}})}+ 
0.227\,{e^{-t/\tau_{_{Co}}}\over (1-\tau/\tau_{_{Co}})}\right] \, {\rm 
erg\, s^{-1}}\,.   
\label{Lumac}
\end{equation}
The light curve after the collision can be approximated by 
\begin{equation}
L(t>t_c)\approx [1-e^{-(t-t_c)^2/t_r^2}]\,[L_c(t-t_c) + L_p(t)]
\label{bolum} 
\end{equation}
where the first term on the RHS is roughly the fraction of the 
volume of the CS shell which is transarent to the radiation 
(approximately  $1/\tau_{opt}\propto t^2$ for $\tau\gg 1$ and 1 for 
for $\tau\lsim 1$), and where we have 
neglected the  spread in arrival times ($\sim R/c$) 
of photons emitted simultaneously in the SN rest frame 
from different points of the photosphere of the CS shell. 
The bolometric luminosity predicted 
by Eqs.~(3) and (4) for a single collision involves five parameters, 
$M({\rm ^{56}Ni})$ that was synthesized in the SN explosion, 
$E_{cm}=\mu\, v^2/2$, and  $t_c$, $ t_{r}$, and $t_d=t_{opt}$,  which 
depend 
on the unknown  mass-loss history ($\dot{M}$, angular distribution  and 
chemical composition) 
of the progenitor star during the years
preceding its SN explosion.
The time-integrated luminosity satisfies
$\int L\,dt=\mu\,v^2/2+ 2.31\times 10^{50}\, M(^{56}{\rm Ni})/M_\odot$,
and can be used to determine the center of mass energy of the 
colliding shells. The product $\eta\, R(t)^2$ can be determined from the 
measured luminosity and black body temperature of the photospherte. 
If after the collision the merged shell overtakes another CS shell, then 
the additional luminosity {\it after} 
the second collision is also described 
by Eq.~(4) but with $t_c(2)$, $ t_{r}(2)$, $E_k(2)$ and $\tau(2)$ 
corresponding to that collision.

\section{Comparison with observations}
In Figures 1-3 we compare the bolometric light-curves of three 
representative SLSNe  and their light-curves as predicted by 
Eqs. (3) and (4).\\
Figure 1 compares the bolometric light-curve of SN1998bw (Galama et al.~ 
1998; Nakamura et al.~ 2001; Fynbo et al.~2000) which produced GRB 980425 
(Pian et al.~2000) and the light-curve predicted by Eqs.~(3) and (4), 
assuming it was powered by a single collision between the SN debris and a 
CS shell ejected sometime before the SN/GRB event and by the decay of 
$^{56}$Ni synthesized in the explosion.  The best fit parameters are 
reported in Table 1. The late-time decay of the bolometric light curve 
could be powered by  $0.06\, M_\odot$ of $^{56}$Ni that was synthesized in the 
explosion, if the merged shell is opaque to $\gamma$ rays .\\
Figure 2 compares the bolometric light-curve of the ultra-luminous SN 
2010gx (Pastorello et al.~2010) and the light-curve predicted by Eqs.~(3) 
and (4) for a collision between the SN ejecta and a massive CS shell. The 
best fit parameters are reported in Table 1. The lack of late-time data 
does not allow determination of the mass of $^{56}$Ni synthesized in the 
explosion.
\\ Figure 3 compares the bolometric light-curve of the ultra-luminous SN 
2006gy (Smith et al.~2008,2010) and the light-curve predicted by Eqs.~(3) 
and (4) for a collision between the SN ejecta and a CS shell,
followed by an encounter  between the merged SN-CS shells with a wind or a 
second CS shell downroad around $\sim$ day 140. The best fit 
parameters are  reported in Table 1. The data do
not allow a reliable determination of the time of the second collision
or of the amount of $^{56}$Ni synthesized in the explosion. 
 
\begin{deluxetable}{cllllll}
\tablewidth{0pt} 
\tablecaption{The best fit values of the parameters 
used in reproducing the measured bolometric light curve 
of several SLSNe. Times are measured in days after the SN explosion.}
\tablehead{
\colhead{SN} & \colhead{$t_c$} & \colhead{$t_{max}(L)-t_c$} & 
\colhead{$t_r $} &\colhead{$t_d$} & \colhead{$M(^{56}{\rm Ni})$}&
$\chi^2/dof$}
\startdata
SN1998bw  & 0.52   &      & 7.52  & 39.7 &  $0.06\,M_\odot$& 0.34 \cr
SN2010gx  &        &23.52 & 41.77 & 12.97 & $<0.10\,M_\odot$    & 1.40\cr
SN2006gy  & 5.54  &   & 60.97 & 35.14 &    &           \cr
           & 147.5 &   & 57.22 & 79.15 &   & 1.00       \\
\hline
\enddata
\end{deluxetable}

\section{Discussion and conclusions} 

Very large photospheric velocities were inferred from the early-time broad 
line spectra of SNe associated with an observed long GRB (SNe-GRB) such 
as SN1998bw, 
SN2003dh, and SN2003lw. Also unusual large quanities of $^{56}$Ni 
synthesized in these explosions were inferred from both their bolometric 
lightcurves and spectra (e.g., Nakamura et al. 2001; Mazzali et al. 2001, 
2006). The unusual large value of the kinetic energy of the SN explosion 
that was inferred from the very large early-time photospheric velocities 
led Iwamoto et al. (1998) and the above authors to conclude that SNe-GRB 
belong to a class of hyper-energetic SNe ("hypernovae"), where the kinetic 
energy of the ejecta is typically $5\times 10^{52}$ ergs, and the 
synthesized mass of $^{56}$Ni is $\sim 0.5\, M_\odot$.

However, the early time-photospheric velocity that was inferred, e.g., 
from the broad lines in the spectrum of SN1998bw (Patat et al. 2000) 
decreased by a factor 4 from $\sim$ 40,000 km s$^{-1}$ to $\sim 10,000$ km 
s$^{-1}$ within the first 30 days after the explosion. If these velocities 
were the bulk motion c.m. velocities of the whole SN ejecta, such a 
deceleration would have required collision with a mass $M\sim 
3\,M_{ej}\sim 30\, M_\odot$ enclosed within $R\sim 5\times 10^{15}$ cm.  
But, a typical ISM baryon density of 1 cm$^{-3}$, yields an ISM mass 
$M\sim ~ 4 \pi\, m_p\, R^3/3 \sim 1\times 10^{-10}\, M_\odot$ within such 
a radius, while a wind environment will have typically only $\dot{M}\, R/V 
\lsim 10^{-3}\, M_\odot$ within $R\sim 5\times 10^{15}$ cm.  Hence, either 
the observed velocity was only of  thin photospheric layer of the SN shell 
which decelerated rapidly by collision with a massive wind/shell 
(while the mean velocity of the SN shell was its $\sim 5000$ km/s)
or the broad absorption lines were due to line broadenig 
e.g., by  Compton scattering inside an
optically thick the SN shell). In both cases the kinetic energy of the 
explosion should have been estimated from the late-time nebular velocity 
of $M_{ej}+ M_{css}$ rather than from modeling the early time photospheric 
velocity (Nakamura et al. 2001; Mazzali et al. 2001, 2006). 
The typical observed expansion
velocities, ~5000 km/s, during the nebular phase 
of SNe associated with observed long duration GRB
imply kinetic energy release of only 
$\sim 2.5\times 
10^{51}\,(M_{ej}+M_{css})/10 M_\odot$ erg typical to ordinary SN 
explosions  and do not support an "hypernovae" origin of SN-GRBs. 

Also the true values of the mass of $^{56}$Ni which were synthesized in 
SN1998bw and other SN-GRBs may be much smaller than inferred, e.g., by 
Nakamura et al.(2001) and Mazzali et al.(2001,2006). These large masses  
were inferred mainly from the peak-luminosity in the 
photospheric phase, while in our model, and in reality, a large part of it 
could be supplied by the collision between the SN shell and the CS shell. 
The inferred mass from the nebular phase is highly model dependent since 
it depends on the fraction of the radioactive energy release 
that is absorbed in the SN shell that depends on the unknown mass of the 
shell and its density distribution as function of time, and on the density 
distribution of $^{56}$Ni within it:
The fraction of the total $\gamma$-ray energy that is absorbed in the SN 
shell is $1-1/\tau_\gamma + e^{-\tau_\gamma}/\tau_\gamma$. For 
nearly-transparent SN shells, $\tau_\gamma\ll 1$, and only a fraction 
$\tau_\gamma/2\ll 1$ of the total $\gamma$-ray energy is deposited in the 
SN shell. Hence, in the models of Mazzali and collaborators, a much larger 
mass of $^{56}$Ni was required in order to power the observed luminosity 
of SNe-GRB in both the photospheric and nebular phases. However, because 
the radius of of the SN shell expands like $R\approx V_{ej}\, t$, 
$\tau_\gamma$ decreases like $t^{-2}$. Neglecting other losses, the 
luminosity powered by the decay of $^{56}$Co must decline then like 
$t^{-2}\, e^{-t/111.4\,{\rm d}}$. The observed bolometric light-curve of 
SN1998bw during the time interval 300-778 day, however, displayed an 
exponential decay consistent with that of $^{56}$Co without the $t^{-2}$ 
modulation. This is possible if either the CS shell is opaque to both the 
$\gamma$-rays and the positrons from the decay of $^{56}$Co, or opaque 
only to the positrons.  In our model the SN shell deposits its radio-isotopes
at the bottom of a CS shell. If the lightcurve of SN1998bw is powered also 
by the SN shell collision with a CS shell, which is nearly opaque to both 
the $\gamma$-rays and the positrons from the decay of $^{56}$Co it implies 
a rather small, $\sim\, 0.06\, M_odot$ (i.e., normal).

SLSNe  are probably stripped-envelope SNe, most of which are SNeIc and 
SNeIIn (e.g., Pastorello et 
al.~2010; Quimby et al.~2011 and references therein). Their luminosity is powered mainly 
by plastic collision between their fast 
ejecta and slowly expanding massive circum-stellar shells formed in 
eruptions during the final stage of their life before their explosion. 
SLSNe may also produce GRBs, but like SNeIc-GRBs, most of them  are 
not observed because they are beamed away from our line of sight 
to the SN explosion.

The CS environments of  core-collapse SNe provide evidence of massive winds and 
ejection episodes of massive shells, probably in thermonuclear eruptions 
preceding their SN explosion.  Although it defies current paradigms of 
stellar evolution theory, perhaps the expulsion of a large fraction of 
the stellar mass by winds and thermonuclear eruptions preceding the SN 
explosion of massive stars make it possible for the energy deposition by 
the shock and the neutrinos from their collapsing core to unbind the 
left-over external mass and impart to it a kinetic energy of the order of 
$E_k\sim $ several $10^{51}$ ergs.

\newpage 
\begin{figure}[]
\centering 
\vspace{-2cm} 
\epsfig{file=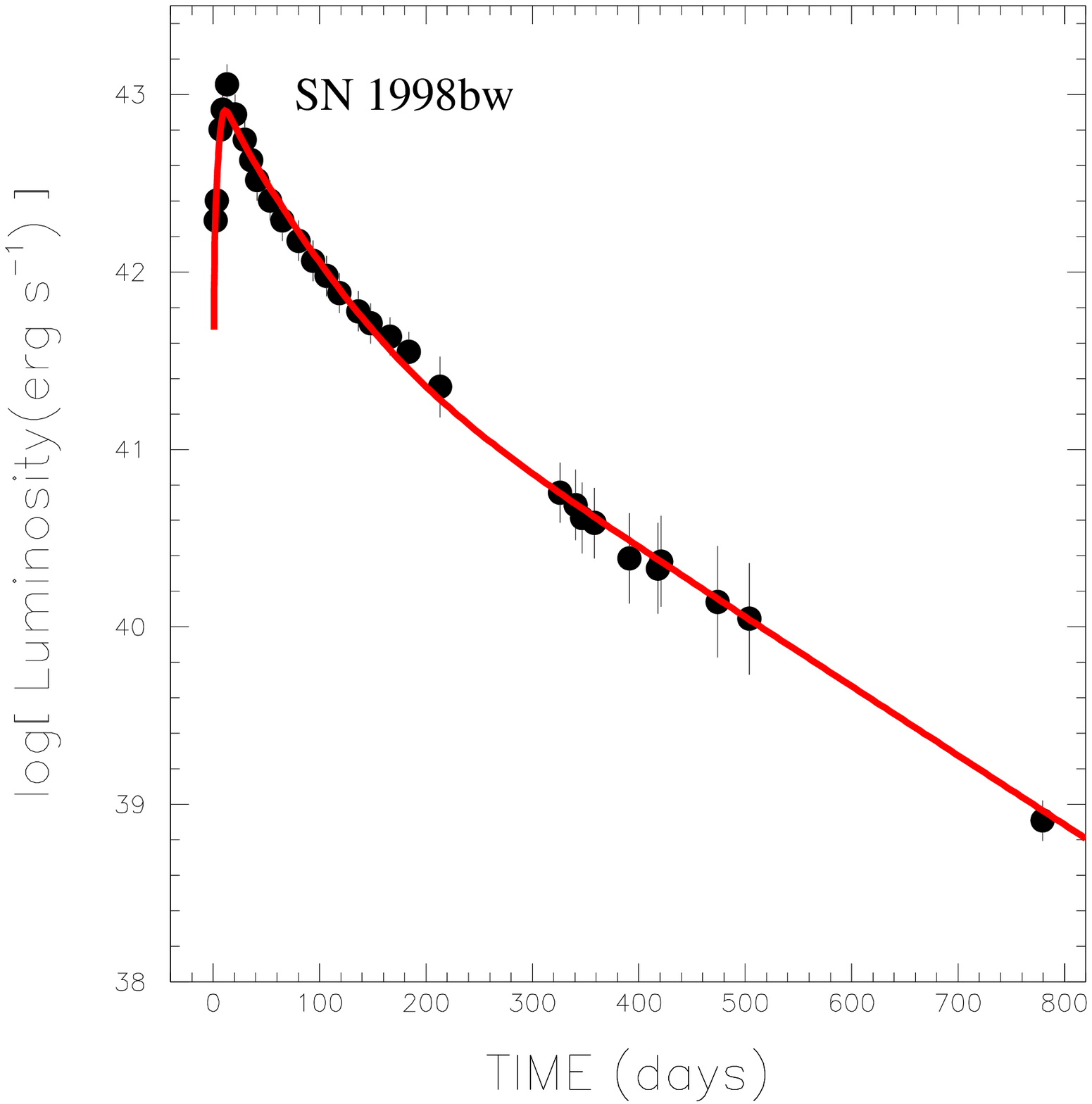,width=12.cm,height=12.cm} 
\caption{ 
Comparison between the bolometric light-curve of SN1998bw 
(Nakamura et al.~2001; Fynbo et al.~2000) and that predicted 
by Eqs.~(3) and (4) assuming it was powered by a plastic collision 
between the SN ejecta and a CS shell/wind and by the decay 
of $^{56}$Ni synthesized in the SN explosion.}   
\label{Fig1}
\end{figure}

\newpage
\begin{figure}[]
\centering
\vspace{-2cm}
\epsfig{file=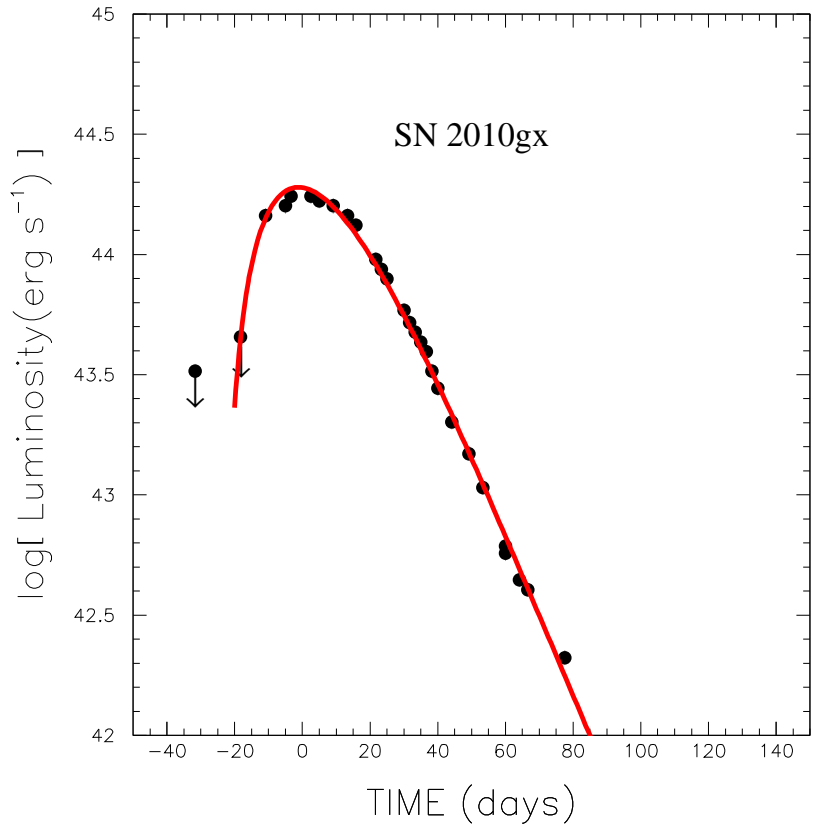,width=12.cm,height=12.cm}
\caption{
Comparison between the bolometric light-curve of SN2010gx 
(Pastorello et al.~2010) and that predicted by Eqs.~(3) and (4)
assuming it was powered by the 
plastic collision between the fast ejecta from the
SN explotion and a much slower massive cs shell ejected
sometime  before the SN explosion.}
\label{Fig2}
\end{figure}

\newpage 
\begin{figure}[]
\centering 
\vspace{-2cm} 
\epsfig{file=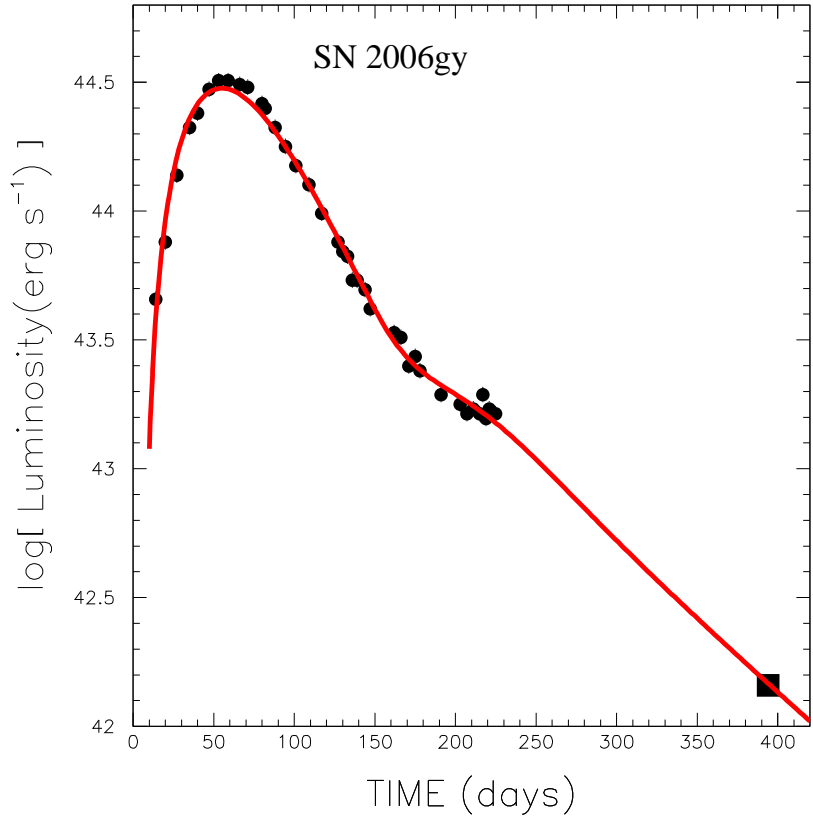,width=12.cm,height=12.cm} 
\caption{ 
Comparison between the oserved bolometric light-curve of SN2006gy 
(Smith et al.~2011)
and that predicted by Eqs.~(3) and (4) 
assuming that it was powered by 
plastic collisions between the fast ejecta of 
SN2006gy and a much slower massive circun-stellar  shells that were 
ejected by its progenitor star in two consecutive eruptions  
sometime  before the  SN  explosion.}
\label{Fig3}
\end{figure}

\begin{thebibliography}


\bibitem[Balberg and Loeb(2011)]{Balberg2011}
Balberg, S. \&  Loeb A. 2011, MNRAS, 414, 1715

\bibitem[Barkat and Rakavy(1992)]{Barkat1967}
Barkat, Z., Rakavy, G., \& Sack, N. 1967, PRL, 18, 379

\bibitem[Campana et al.(2006)]{Campana2006}
Campana, S., et al. 2006, Nature, 442, 1008

\bibitem[Chatzopoulos et al.(2011)]{Chatzopoulos2011}
Chatzopoulos, E., et al. 2011, ApJ, 729, 143

\bibitem[Chatzopoulos et al.(2012)]{Chatzopoulos2012}
Chatzopoulos, E., et al. 2012, ApJ, 746, 121

\bibitem[Chevalier and Irwin(2011)]{Chevalier2011} 
Chevalier, R. A. \& Irwin C. M. 2011, ApJ, 729, L6

\bibitem[Chevalier and Irwin(2012)]{Chevalier2012} 
Chevalier, R. A. \& Irwin C. M. 2012, ApJ, 747, L17

\bibitem[Chugai(1990)]{Chugai1990}
Chugai, N. N., 1990, Sov. Astr. Lett., 16, 457

\bibitem[Chugai(1992)]{Chugai1992}
Chugai, N. N., 1992, Sov. Astr. 36, 63


\bibitem[Cusumano et al.(2006)]{Cusumano2006}
Cusumano, G. et al. 2006, GCN Circ. 4775

\bibitem[Dado et al.(2009)]{Dado2009}
Dado, S., Dar, A. \& De R\'ujula, A. 2009, ApJ, 693, 311

\bibitem[Dado and Dar(2012)]{Dado2012}
Dado, S. \&  Dar, A. 2012, preprint, arXiv:1203.5886 

\bibitem[Dar and De R\'ujula(2000)]{Dar2000}
Dar, A., \& De R\'ujula, A. 2000, arXiv:astro-ph/0008474

\bibitem[Dar and De R\'ujula(2004)]{Dar2004}
Dar, A. \& De R\'ujula, A.~2004, Phys. Rep. 405, 203

\bibitem[Falk and Arnett(1973)]{Falk1973}
Falk, S. W. \& Arnett, W. D. 1973, ApJ, 180, L65

\bibitem[Falk and Arnett(1977)]{Falk1977}
Falk, S. W. \& Arnett, W. D. 1977, ApJS, 33, 515

\bibitem[Fynbo et al.(2000)]{Fynbo2000}
Fynbo, J. U., et al. 2000, ApJ, 542, L89

\bibitem[Galama et al.(1998)]{Galama1998}
Galama, T. J., et al. 1998, Nature, 395, 670

\bibitem[Gal-Yam et al.(2009)]{Gal-Yam2009}
Gal-Yam, A.,  et al. 2009, Nature, 462, 624

\bibitem[Ginzburg and Balberg(2012)]{Ginzburg2012}
Ginzburg, S. \&  Balberg, S. 2012, preprint  arXiv:1205.3455

\bibitem[Grassberg et al.(1971)]{Grassberg1971}
Grassberg, E. K., Inshennik, V. S. \& Nadyozhin, D. K. 1971,
Ap\&SS, 10, 28

\bibitem[Heger and Woosley(2002)]{Heger2002}
Heger, A., \& Woosley, S. E. 2002, ApJ, 567, 532

\bibitem[Hjorth et al.(2003)]{Hjorth2003} 
Hjorth, J., et al. 2003, Nature, 423, 847 

\bibitem[Iwamoto et al.(1998)]{Iwamoto1998}
Iwamoto, K., et al. 1998, Nature, 395, 672

\bibitem[Kodros et al.(2010)]{Kodros2010}
Kodros, J., et al. 2010, CBET 2461

\bibitem[Knop et al.(1999)]{Knop1999}
Knop, R., et al. 1999,  IAU Circ. 7128 

\bibitem[Mazzali et al.(2001)]{Mazalli2001}
Mazzali, P. A., et al. 2001, ApJ, 559, 1047)

\bibitem[Mazzali et al.(2006)]{Mazalli2006}
Mazzali, P. A., et al.  2006, ApJ, 645, 1323

\bibitem[Moriya et al.(2012)]{Moriya2012}
Moriya, T., et al. 2012, preprint arXiv:1204.6109

\bibitem[Nakamura et al.(2001)]{2001}
Nakamura, T.,  et al.  2001, ApJ, 550, 991

\bibitem[Ofek et al.(2007)]{Ofek2007}
Ofek, E. O., et al. 2007, ApJ, 659, L13

\bibitem[Ofek et al.(2012)]{Ofek2012}
Ofek, E. O., et al. 2012,  preprint arXiv:1206.0748

\bibitem[Pastorello  et al.(2010)]{Pastorello2010}
Pastorello, A., et al. 2010, ApJ, 724, L1


\bibitem[Pian et al.(2000)]{Pian2000}
Pian, E.,  et al. 2000, ApJ, 536, 778

\bibitem[Pian et al., 2006]{Pian2006}
Pian, E.,  et al. 2006, Nature, 442, 1011

\bibitem[Quimby et al.(2007)]{Quimby2007}
Quimby, R. M., et al. 2007, ApJ, 668, L99

\bibitem[Quimby et al.(2011)]{Quimby2011}
Quimby, R. M., et al. 2011, Nature, 474, 487

\bibitem[Rakavy and Shaviv(1967)]{Rakavy1967}
Rakavy, G. \& Shaviv, G. 1967, ApJ, 148, 803

\bibitem[Rest et al.(2011)]{Rest2011}
Rest, A., et al. 2011, ApJ, 729,  88

\bibitem[Smith and McCray(2007)]{Smith2007}
Smith, N. and McCray, R. 2007, ApJ, 671, L17 

\bibitem[Smith et al.(2008)]{Smith2008}
Smith, N., et al. 2008, ApJ, 686, 485

\bibitem[Smith et al.(2010)]{Smith2010}
Smith, N., et al. 2010, ApJ, 709, 856

\bibitem[Soffitta et al.(1998)]{Soffitta1998}
Soffitta, P., et al. 1998, IAU Circ. 6884

\bibitem[Sollerman et al.(2006)]{Sollerman2006}
Sollerman, J., et al. 2006, A\&A, 454, 503 

\bibitem[Stanek et al.(2003)]{Stanek2003}
Stanek, K. Z., et al. 2003, ApJ, 591, L17 

\bibitem[Stoll et al.(2011)]{Stoll2011}
Stoll,  R.,  et al. 2011, ApJ, 730, 34

\bibitem[Vanderspek, et al.(2003)]{Vanderspek2003}
Vanderspek, R., et al. 2003, GCN Circ. 1997

\bibitem[Vinko et al.(2010)]{Vinko2010}
Vinko, J., et al. 2010, CBET 2476

\bibitem[Waldman(2008)]{Waldman2008}
Waldman, R. 2008, ApJ, 685, 1103

\bibitem[Yoshida and Umeda(2011)]{Yoshida2011}
Yoshida, T., \& Umeda, H. 2011, MNRAS, 412, L78
\end{thebibliography}
\end{document}